\documentclass[preprint,aip,graphicx,onecolumn]{revtex4}
\usepackage{graphicx}
\usepackage{color}
\usepackage{ulem}
\begin{document}


\title{Thermodynamic consistency in variable-level
coarse-graining of polymeric liquids}

\author{A. J. Clark}
\author{J. McCarty}
\author{I. Y. Lyubimov}
\author{M. G. Guenza\footnote{Author to whom correspondence should be addressed. Electronic mail: mguenza@uoregon.edu}}
\affiliation{Department of Chemistry and Institute of Theoretical Science
University of Oregon, Eugene, OR 97403, USA}



\begin{abstract}
Numerically optimized reduced descriptions of macromolecular liquids often present thermodynamic inconsistency with atomistic level descriptions even if the total correlation function, i.e. the structure, appears to be in agreement.
An analytical expression for the effective potential between a pair of coarse-grained units is derived starting from the first-principles Ornstein-Zernike equation, for a polymer liquid where each chain is represented as a collection of interpenetrating blobs, with a variable number of blobs, $n_b$, of size $N_b$. The potential is characterized by a long tail, slowly decaying with characteristic scaling exponent of $N_b^{1/4}$. This general result applies to any coarse-grained model of polymer melts with units larger than the persistence length,  highlighting the importance of the long, repulsive, potential tail for the model to correctly predict  both structural and thermodynamic properties of the macromolecular liquid.
  
  
\end{abstract}


\maketitle 


Relevant structural and dynamical properties of polymer liquids take place over a large range  of length (and time) scales.  The computational requirements often render it impractical or impossible to fully investigate such systems with atomistic level simulations. To overcome these limitations, many descriptions of polymeric liquids with reduced internal degrees of freedom, or coarse-grained (CG) descriptions, have been proposed. The tremendous interest that CG methods have generated is due to their capability of speeding up simulations, probing systems on larger length scales than conventional atomistic simulations.\cite{molinero,dePablo,papoian,mullplathe,kremer,klein,kremer2,Pierleoni} 

The basis for any coarse grained description  is the specification of effective interaction potential energies or forces between coarse grained units. Despite the growing interest in CG methods and their proven successes, most CG models are still limited in their potential application because of the empirical character of the effective interaction potentials upon which they rely. While a few CG approaches have been formally derived,\cite{Voth,Noid,Hansen,ours} most, for example the Iterative Boltzman Inversion (IBI) procedure where the mesoscale potential is optimized to reproduce the total correlation functions, and sometimes secondarily the pressure,\cite{mullplathe} rely on numerical optimization of their CG parameters through comparison with the related atomistic simulations or with experimental data. Unfortunately, numerically optimized CG potentials are in principle neither transferable to different systems, nor to the same system in different thermodynamic conditions, and they can not assure 
consistency for properties 
different from the ones against which their parameters have been optimized\cite{Louis}. This limits dramatically their generality and convenience.


In this letter we present an analytical formalism for the
potential describing the interaction between CG units, for  a model of a polymer liquid where each molecule is represented as a chain of interpenetrating spheres  with variable size and number, in variable thermodynamic conditions: we show that the CG model presented and the related analytical potential, ensure structural and thermodynamic consistency with the atomistic description.
The model is general, as the  
potential is fully transferable and it is applicable to liquids of polymers with different molecular structure. The analytical form of the potential allows for the characterization of some general features of how thermodynamic properties and structure depend on the shape of the intermolecular potential between CG units. 

In a polymeric liquid of monomer density, $\rho_m$, and number of polymers $\mathit{n}$, the size of the polymeric chain is defined by its radius-of-gyration, $R_g^2=Nl^2/6$, with $l$ the effective segment length between the center-of-mass (coms) of two monomers, and $N$ the number of monomers in a chain. Here we are concerned about liquids of polyethylene for which $l=0.152 \ nm$. 
Each polymer is represented as a chain of superimposing soft blobs, with $N_b$ the number of monomers in one blob, $n_b=N/N_b$ the number of blobs in one chain and blob size characterized by, $R_{gb}=R_g/\sqrt{n_b}$, 
The correlation functions that describe the static structure of the blob CG model have been derived from the formal solution of a generalized 
Ornstein-Zernike equation where monomeric sites are assumed to be real sites, and CG sites are assumed to be fictitious sites, in an extension of Krakoviack's et al. original approach,\cite{krakov,krakov1} and including the thread model polymer reference interaction site model (PRISM) monomeric description.\cite{PRISM,Clark} By assuming a Gaussian distribution of the monomers inside a blob and monomer interactions much shorter ranged than the size of the blob, which are reasonable approximation for subchains  with $N_b$  larger than the persistence length (for polyethylene $N_b \ge 30$), an analytical blob-blob total correlation function,  $\hat{h}^{bb}(r )$, was derived, and shown to be in quantitative agreement with the total correlation function from atomistic simulations.\cite{Clark} 

Starting from $\hat{h}^{bb}(r )$ and given that the monomer direct correlation function, $c^{mm}(r)$, has a range shorter than the size of the CG unit, an analytical solution of the CG potential is derived by approximating the Fourier transform of $c^{mm}(r)$ by its zero wavevector value, $c_0=4 \pi \int_0^{\infty}r^{2}c^{mm}(r)$. 
The effective direct correlation between blobs is given in Fourier space by  
\begin{equation}
 \hat{c}^{bb}(k) = -\frac{N_b \Gamma_b}{\rho_m} \frac{(\hat{\Omega}_{av}^{bm}(k)/\hat{\Omega}_{av}^{bb}(k))^2}{1+n_b \Gamma_b [\hat{\Omega}^{mm}(k)-(\hat{\Omega}^{bm}_{av}(k))^2/\hat{\Omega}_{av}^{bb}(k)]} \ , 
\label{cofk}
\end{equation}
where $\Gamma_b= - N_b \rho_m c_0$ where $c_0 < 0$, and $\hat{\Omega}_{av}^{bm}(k)$ is the intramolecular correlation function between the coms of a blob and a monomer, averaged over the chain and normalized such that its value at $k=0$ is equal to one.
The sharp peak around $k=0$ leads to the enhanced sensitivity of the function to the quality of the approximations used in the chain model, and made numerical solutions of the potential necessary in our previous work.
The total correlation function in this blob description gives isothermal compressibility, $\kappa_T =[ k_B T \rho_{m}(1+\Gamma_b)/N ]^{-1}$, with $k_B$ the Boltzmann constant and $T$ the temperature,  which is consistent with the compressibility in the atomistic description\cite{PRISM}. Intra- and inter-molecular pair distribution functions and the structure factor, are consistent with their atomistic counterpart  for distances $r > R_{gb}$ and wave vectors  $k < 2 \pi /R_{gb}$.


Assuming that a coarse grained unit contains a number of monomers sufficiently large  to follow a Gaussian space distribution, and that the density and the interaction strength are large enough that the product $\Gamma_b >> 1$, the contribution to the inverse transform integral for large wavevectors ($k >> 1/R_{gb}$) is negligible for $r>R_{gb}$. In this limit, the direct correlation for $k<<1/R_{gb}$ has the simple rational function limiting form $c^{bb}(k) \propto 1 / (1+\Gamma R_{gb}^4 k^4)$ as $\Gamma_b \to \infty$. Approximating the effective direct correlation by this form for all wavevectors introduces very little error, allowing for a simple approximation for the functional form in real space.  
The accuracy at intermediate $\Gamma$ values can be improved  by taking this as the zeroth order term of an asymptotic expansion in $1 / \sqrt{\Gamma_b}$ about $\Gamma_b \to \infty$.  

The effective potential is then derived by applying the approximation $V^{bb}(r) \approx -k_B T [c^{bb}(r)-h^{bb}(r)+ln(1+h^{bb}(r))] \approx -k_B T c^{bb}(r)$, the limiting form of the Hyper-Netted Chain approximation valid when $|h^{bb}(r)|<<1$ everywhere.  This approximation holds for soft potentials in the limit of high densities and long chains of interest here.\cite{HansenMcDonald} For $r>R_{gb}$, the intermolecular blob potential, which is the needed input for the mesoscale simulations of the coarse-grained polymer liquid, is given by  

\begin{equation}
\begin{array}{rcl}
 V^{bb}(r) &\approx& k_B T \big[ \left(\frac{45 \sqrt{2} N_b \Gamma_b^{1/4}}{8 \pi \sqrt{3} \sqrt[4]{5} \rho_m R_{gb}^3}\right)\frac{sin(Qr)}{Qr}e^{-Qr} \\
&&-\left( \frac{\sqrt{5}N_b}{672\pi\rho_m \Gamma_b^{1/4} R_{gb}^3} \right) \big[(13Q_{rs}^3(Qr-4))cos(Qr)
\\ &&+\left(\frac{945+13Q_{rs}^4}{\Gamma_b^{1/4}}\right)rsin(Qr)+\frac{945r}{\Gamma_b^{1/4}} cos(Qr)\big] \frac{e^{-Qr}}{Qr} \big] \ ,
\end{array}
\label{Vccr-apprx}
\end{equation}

\noindent where $Q=5^{1/4}\sqrt{3/2}/(\Gamma_b)^{1/4}$, $Q_{rs}=5^{1/4}\sqrt{3/2}$, and $r$ is in units of the blob radius-of-gyration, $R_{gb}$.
When $n_b=1$ Eq.(\ref{Vccr-apprx}) represents the effective potential between the center-of-mass of two polymers in a melt.  

Because the potential is formally expressed as a function of the molecular and thermodynamic parameters, Eq.(\ref{Vccr-apprx}) is in principle general and applicable to polymer melts in different thermodynamic conditions and with diverse macromolecular structures.

The range of the effective potentials between CG units scales beyond the effective blob radius of gyration, $R_{gb}$, and decays with the number of monomers per blob as $N_b^{1/4}$. 
%
The observed scaling behavior can be explained by considering that in a liquid the total effective correlation between two sites (atomistic or CG), and its related potential, can be regarded as ``propagating'' through sequences of direct pair interactions following the Ornstein-Zernike integral equation theory. These many-body contributions to the pair interaction are not simply additive, and once mapped into the OZ pair interaction, they result in a slowly decaying tail.  Because of the Gaussian statistics that applies to the structure of long polymeric chains, the interaction between any intermolecular pair of blobs statistically propagates in the shared volume as a random walk following the random path of effective CG sites. Given that in the relevant volume $V_b \propto R_{gb}^3\propto  N_b^{3/2} l^3$, the number of effective CG sites is of the order of $n'_b\propto \rho_m R_{gb}^3/N_b \propto N_b^{1/2} \rho_m l^3$, which leads to the observed scaling of $N_b^{1/4}$.

From the analytical form of the effective potential, Eq.(\ref{Vccr-apprx}), the pressure of the system can be calculated explicitly.  Specifically in the high density, long chain limit, the pressure calculated in the virial route reduces to the simple expression 
\begin{equation}
 \frac{P}{\rho_c k_B T} \approx 1-\frac{N c_0 \rho_m}{2} \ .
 \label{virial}
\end{equation}
where $\rho_c=\rho_m/N$ is the chain density. Eq.(\ref{virial}) is in agreement with the monomer level description,\cite{PRISM} and does not depend on the level of coarse-graining of the model selected. It should be noted that because both the pressure and the compressibility are consistent with monomer level PRISM, the equations of state predicted from the virial and compressibility routes will show the same small inconsistencies between the routes, which are inherent to the approximations in integral equation theory.\cite{PRISM,HansenMcDonald}  The important point is that the pressure and compressibility are left unchanged from the monomer level description by the coarse graining procedure on all levels.

To test the self consistency and the predictions of the effective potentials we performed molecular dynamics (MD) simulations of polymer melts using the LAMMPS simulation package,\cite{LAMMPS} in parallel on the SDSC Trestles cluster accessed through the XSEDE project. Simulations were performed both at the atomistic level, by adopting the traditional united atom model (UA-MD) with the established set of UA potentials (intermolecular Lennard-Jones,\cite{TRAPPE, UAmodel} harmonic bonds and angle potentials) 
and at the coarse-grained mesoscale level (MS-MD) where each chain was represented as a collection of soft colloidal particles. 
MS-MD simulations for the $n_b=1$ case have been described previously\cite{McCarty-jpcb,blends,Lyubimov}. To extend the model to cases where $n_b>1$, the effective bond between adjacent blobs was taken to be $V_{bond}=3 k_B T r^2 /(8 R_{gb}^2) + V^{bb}(r) + k_B T ln(1+h^{bb}(r))$, which enforces the correct distributions between adjacent blobs\cite{Clark}, and an angle potential between sequential triples on each chain which likewise enforces the correct angular probability distribution.\cite{Lasos} Blobs more than two apart on each chain were taken to interact via the intermolecular pair potential $V^{bb}(r)$ of Eq.(\ref{Vccr-apprx}). All Mesoscale simulations used here are performed in the NVE (microcanonical) ensemble.
Values of the $c_0$ parameter, entering $\Gamma_b$ in Eq.(\ref{Vccr-apprx}) and the MS-MD, were taken from the UA simulations for short chains. For systems with large $N$, which were too slow to relax to be accessible through UA-MD,   the $c_0$ parameter was extrapolated from the available UA data at small $N$ at the same monomer density, using the form predicted from monomer level numerical PRISM theory, $c_0=a+b/N$, with $a$ and $b$ optimized parameters. The values of $c_0$ obtained from this procedure were found to be generally consistent with calculations using long-established numerical monomer level PRISM models when an attractive part to the monomer potential is included.\cite{PRISM}  

\begin{figure}
$\begin{array}{c}
 \includegraphics[width=3.0in]{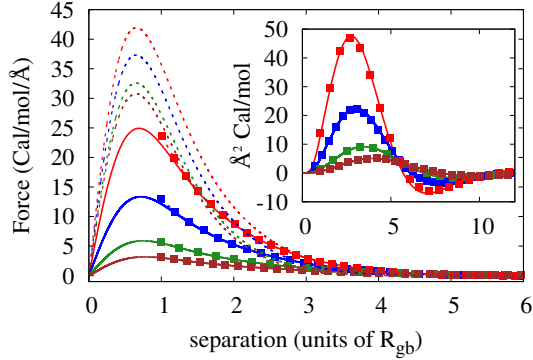} \\
\end{array}$
\caption{ Force curves for the $n_b=1$ model (solid lines) and the approximated Eq.(\ref{Vccr-apprx}) (squares) for (top to bottom) $N=100$ (red), $N=200$ (blue), $N=500$ (green), and $N=1000$ (brown), along with force curves for models with a fixed blob length, $N_b=50$, and increasing numbers of blob (dashed lines): $n_b=2$ (red), $n_b=4$ (blue),  $n_b=10$ (green), and $n_b=20$ (brown).  Inset: Normalized Virial density ($N^{-2} r^3 F^{bb}(r)$) for the numerical forces  and their analytical forms for $n_b=1$.  All systems are at 400K and density $\rho_m=0.03355 \AA^{-3}$.}
\label{press-vs-N1}
\end{figure}

\begin{figure}
$\begin{array}{c}
\includegraphics[width=3.0in]{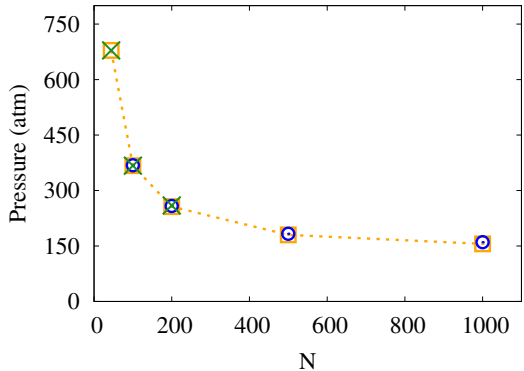} 
 \end{array}$
\caption{Pressure measured from simulations with input CG forces from Figure \ref{press-vs-N1}, where $n_b=1$ (orange squares connected by dashed guide lines), and $N_b=50$ with $n_b> 1$ (open blue circles). Also depicted are UA-MD simulations (green X symbols) for systems which relax fast enough for UA-MD to be feasible ($N \leq 200$).  Data are collected for increasing degree of polymerization $N$, at the same values of $N$ as Fig.(\ref{press-vs-N1}).  All simulations were performed at 400K and density $\rho_m=0.03355 \AA^{-3}$.}
\label{press-vs-N3}
\end{figure}

Figure \ref{press-vs-N1}  displays the effective intermolecular force ($-\partial V^{bb}/\partial r$) between blobs, comparing the numerical solution of the potential with the approximate expression, Eq.(\ref{Vccr-apprx}). The approximate expression represents correctly the force in real space going from the peak to the long tail, which is the essential information needed for the correct calculation of thermodynamic properties. The inset in Figure \ref{press-vs-N1} shows the normalized Virial density ($N^{-2} r^3 F^{bb}(r)$), whose integral is proportional to the pressure, and which can be seen to be dominated by the tail region of the potential ($r > R_{gb}$). The peak of the force decreases with increasing blob length, $N_b$, at constant density, $\rho_m$, but its range increases in such a way that the Virial integral ultimately reaches a plateau, which corresponds to the leveling off of the pressure, as shown in Figure \ref{press-vs-N3}. The inset of Figure \ref{press-vs-N1} also shows that the effective 
potential has a small attractive part at long range: such an attractive component is a necessary condition for a stable liquid to form, as a gas phase would be the state of lowest free energy at any temperature for a system of purely repulsive particles with no additional constraints. This attractive contribution is partially of entropic origin and partially due to the attractive component of the inter-monomer potential.
Figure \ref{press-vs-N3} shows that all simulations  of the same system, independent of the level of coarse graining, generate not only consistent structure, as seen before, but also consistent pressure.


%
%
%
%
%
%
%
\begin{figure}
$\begin{array}{c}
\includegraphics[width=3.0in]{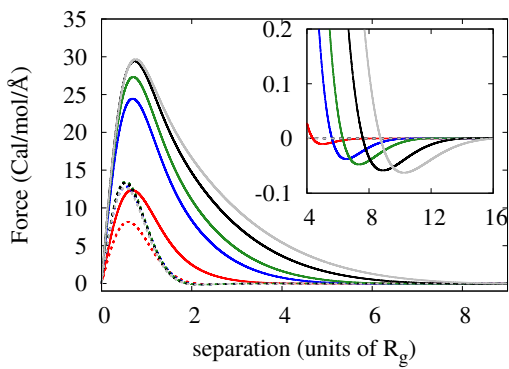} \\
 \end{array}$
\caption{ Numerical force curves for $N=100$, $n_b=1$ models of PE with increasing densities (bottom to top) $0.03226 \AA^{-3}$ (red), $0.03355 \AA^{-3}$ (blue), $0.03441 \AA^{-3}$ (green), $0.03656 \AA^{-3}$ (black), and $0.03871 \AA^{-3}$ (gray), along with corresponding mean force curves (dashed lines). Insert: detail of the attractive contribution to the force (densities increase left to right).  All systems are at 400K }
\label{press-vs-rho1}
\end{figure}

\begin{figure}
$\begin{array}{c}
\includegraphics[width=3.0in]{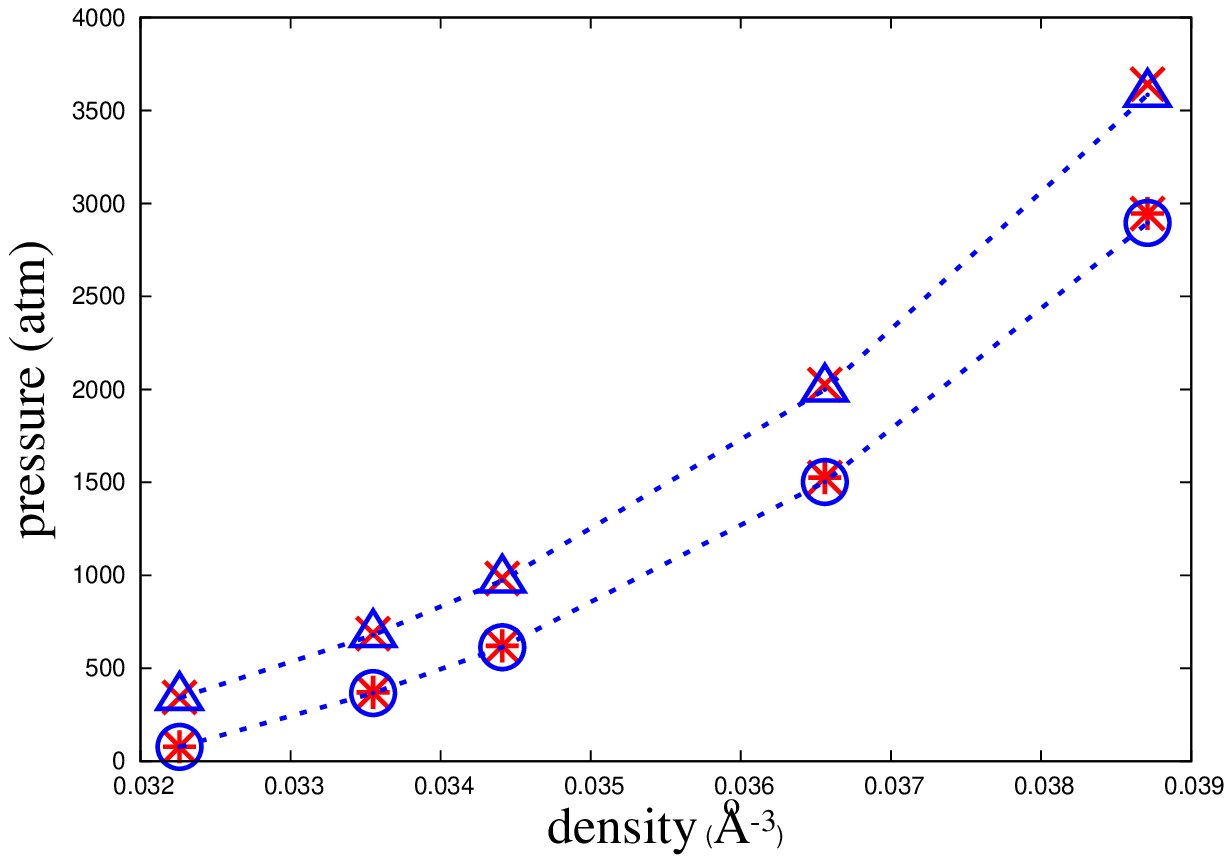}
\end{array}$
\caption{Pressures measured in MS-MD simulations versus density, for $N=44$ (triangles) and $N=100$ (circles), performed with effective potentials where $c_0$ is determined from UA-MD simulations. 
Also reported are values from the corresponding UA-MD simulations (crosses for $N=44$ and stars for $N=100$).  All simulations were performed at 400K
Connecting lines are added as a guide to the eye. }
\label{press-vs-rho2}
\end{figure}

The scaling of the potential is a property of the representation; increasing the number of monomers in each coarse grained unit on each chain, with a fixed system density and monomer interaction, results in an effective potential that grows long-ranged because it captures the average effect of the many-polymer correlations mapped into the effective pair interactions. This is a consequence of the fractal dimension of polymers, which is of order two, as blob volume increases with $N_b$ faster than the chain can fill it, leading to an increasing with $N_b$ of the level of chain interpenetration.
If however, melts of increasingly long chains at the same density are represented by increasing numbers of soft blobs of fixed $N_b$, the range of the potential between blobs limits to a fixed value with increasing chain length as the direct correlation parameter $c_0$ approaches its limit.

Analysis of these results shows that the range of the effective potential is highly density dependent, both explicitly and indirectly through the direct correlation parameter, $c_0$, as shown in Figure \ref{press-vs-rho1}. Also reported for comparison is the force generated from the potential of mean force, which is a very poor representation of the ``real'' force at high densities.

Finally Figure \ref{press-vs-rho2} compares predictions of pressure as a function of density for two samples with increasing chain length, i.e. $N=44$ and $N=100$.  Data from MS-MD of the CG soft-blob representation show excellent agreement with data from UA-MD in the range where UA-MD were performed. The agreement appears to be independent of the level of CG representation that is adopted, as soft-spheres and chains of soft-blobs have consistent pressure across the different levels of coarse graining. The calculations are dominated by the presence of the long-ranged tail of the potential, further validating the proposed CG description.  The range of separations over which the effective potential must represent the average of many-polymer effects increases dramatically with density.  While all levels of representation accurately reflect whole-system thermodynamic averages and structural pair correlations, a coarse grained description may be unable to resolve processes below the length scale of the potential 
tail, due to this averaging effect.

The difference between the scaling predictions for the effective potential and the potential of mean force  also has very important implications for theories of polymer melts.
The strength of the potential of mean force, $w(r)=- k_BT ln (1+h^{bb}(r ))$, is found to scale for long chains, or large blobs,
and high densities as $1/(\rho_m \sqrt{N})$ and its range as $N^{1/2}$ 
with $c_0$ becoming irrelevant and $w(r )$ vanishing in the infinite chain limit.  Use of the $w(r )$ in thermodynamic relations therefore results in vanishing energies and pressures in the long chain/high density limit, which would imply the irrelevance of intermolecular interactions and allow reduction of the system to a single chain problem.  
While the effective pair potential at contact, $V^{bb}(0 )$, still vanishes in the infinite chain limit, its tail increases in such a way that the Virial integral does not vanish.  Likewise the average effective energy in the system also does not vanish, and furthermore both depend on the monomer level direct correlation, and therefore on the monomer interaction potential.  This implies that even in the infinite chain limit, intermolecular potentials do not become irrelevant, and therefore cannot generally be neglected, as it is conventionally done in  the description of polymer melt dynamics.\cite{DoiEdw}

In this letter we have presented an analytical characterization of the soft pair potentials that arise in high level coarse grained descriptions of interpenetrating polymer melts, i.e. for soft-sphere and chains of soft-blobs coarse graining mappings of polymer melts. The form of the potential and the pressure are obtained directly in terms of parameters from the monomer level theory, without a need for phenomenological correction forms in the potential and numerical optimization procedures. The effective potentials for these CG models show characteristic long-ranged ``tails'' that scale non-trivially with chain length, density and monomer interaction strength. This family of CG models guarantees the consistency of the structure and thermodynamics of the macromolecular liquid in any level of soft representation, which is relevant in the modeling of complex polymeric liquids, as well as in the design of reliable multiscale modeling approaches to capture relevant phenomena that may occur on many different 
length scales.  Dynamical properties will be accelerated in the coarse-grained representation, and will have to be properly rescaled to reconstruct the realistic dynamics.\cite{Lyubimov}
 
We acknowledge support from the National Science Foundation through grant DMR-0804145.
Computational resources were provided by Trestles through the XSEDE project supported by NSF. 
This research was  supported in part by the National Science Foundation under Grant No. PHY11-25915.

\end{document}